# Modeling metallic island coalescence stress via adhesive contact between surfaces

Steven C. Seel[1], Jeffrey J. Hoyt[2], Edmund B. Webb III[2], and Jonathan A. Zimmerman[3]

[1]*Sandia National Laboratories, Surface and Interface Sciences, MS 1415, Albuquerque, NM 87185*

[2]*Sandia National Laboratories, Computational Materials Science and Engineering, Albuquerque, NM 87185*

[3]*Sandia National Laboratories, Science-Based Materials Modeling, Livermore, CA 94551*

Tensile stress generation associated with island coalescence is almost universally observed in thin films that grow via the Volmer-Weber mode. The commonly accepted mechanism for the origin of this tensile stress is a process driven by the reduction in surface energy at the expense of the strain energy associated with the deformation of coalescing islands during grain boundary formation. In the present work, we have performed molecular statics calculations using an embedded atom interatomic potential to obtain a functional form of the interfacial energy vs distance between two closely spaced free surfaces. The sum of interfacial energy plus strain energy provides a measure of the total system energy as a function of island separation. Depending on the initial separation between islands, we find that in cases where coalescence is thermodynamically favored, gap closure can occur either spontaneously or be kinetically limited due to an energetic barrier. Atomistic simulations of island coalescence using conjugate gradient energy minimization calculations agree well with the predicted stress as a function of island size from our model of spontaneous coalescence. Molecular dynamics simulations of island coalescence demonstrate that only modest barriers to coalescence can be overcome at room temperature. A comparison with thermally activated coalescence results at room temperature reveals that existing coalescence models significantly overestimate the magnitude of the stress resulting from island coalescence.

# I. INTRODUCTION

Stresses generated during thin film growth strongly influence component lifetime and performance in applications ranging from microelectronics to mechanical coatings and microelectromechanical systems. These residual stresses can result in failure due to film delamination, cracking at interfaces, and hillock formation. In contrast to their deleterious effects, thin film stresses can also drive strain mediated self assembly of nanostructures such as quantum dots. However, the intrinsic connections between an evolving thin film morphology during growth and the corresponding stress generation mechanisms are still a matter of debate.[1,2]

For films that grow via the Volmer-Weber mode such as metals deposited on oxides, crystallites of critical size nucleate on the substrate surface as isolated islands. With continued deposition, the growing islands impinge and coalesce to eventually form a continuous polycrystalline film. Transmission electron microscopy observations coupled with stress measurements indicate that tensile stress generation during the early stages of film growth is associated with the process of island coalescence.[3–5] Hoffman postulated that during the island impingement stage of growth, neighboring islands will stretch towards each other and coalesce in order to reduce surface energy at the expense of an associated strain energy.[6,7]

Although Hoffman suggested that tensile stress generation is driven by a reduction in surface energy during island coalescence, he did not use this idea to estimate the associated stress. Instead he assumed that as atoms are deposited on an island surface near the point of impingement, they are more likely to arrive in the attractive region of the asymmetric potential well describing atomic interactions thereby resulting in a net tensile attraction between coalesced islands.[6,7] Hoffman interpreted this process as a "constrained relaxation" due to local atomic rearrangement within the grain boundary and not as a uniform stretching of the islands. The resulting "distortion" $\Delta$ of the boundary can be estimated to be slightly less than 1 Å (independent of island size and surface energy) so that the associated average biaxial tensile stress in the film is:

$$\sigma = M \frac{\Delta}{w}, \qquad (1)$$

where $M = E/(1-\nu)$ is the biaxial modulus of the film with Young's modulus $E$ and Poisson ratio $\nu$, and $w$ is the island diameter.[7,8]

Nix and Clemens (NC) were the first to reinterpret Hoffman's original argument and calculate the coalescence stress resulting from the balance between the energy increase due to *uniformly* stretching the islands and the energy decrease due to the elimination of the free surfaces.[9] NC modeled the coalescence of hexagonal islands with vertical side faces, while others have subsequently considered the simpler geometry of an array of square islands[10,11] as shown in Fig. 1(a). Consider a periodic array of square islands of lateral dimension $w$ and height $h$ on a thick substrate. The lateral gap $\alpha$ between neighboring islands is imagined to decrease as a consequence of island growth. At some critical gap size, the islands strain equibiaxially by an amount $\alpha/w$ to eliminate two free surfaces of energy $2\gamma_s$ for every new interface of energy $\gamma_b$. Ignoring traction of the islands with the underlying substrate, the resulting increase in elastic strain energy for each island is $\Delta E_\varepsilon = M(\alpha/w)^2(hw^2)$, while the corresponding change in the surface and interface energy is $\Delta E_\gamma = -4wh(\gamma_s - \frac{1}{2}\gamma_b)$, where each new interface is shared between two islands. The critical gap size $\alpha_{crit}$ for which these two energetic contributions are balanced, i.e., $\Delta E_\varepsilon + \Delta E_\gamma = 0$, is given by:

$$\alpha_{crit} = 2\sqrt{\Delta\gamma \cdot w / M} , \qquad (2)$$

where $\Delta\gamma = (\gamma_s - \frac{1}{2}\gamma_b) > 0$, which is a necessary condition for coalescence to occur. Using Eq. (1) and substituting $\Delta = \alpha_{crit}$, the corresponding biaxial tensile stress is:

$$\sigma = 2\sqrt{\frac{\Delta\gamma \cdot M}{w}} . \qquad (3)$$

Equations (2) and (3) were first derived by NC but in deference to the original motivation for the calculation, we will refer to these results as the Hoffman model. With reasonable values for $M = 100$ GPa (assuming $E = 67$ GPa and $\nu = 0.33$), $\Delta\gamma$ 1 J/m$^2$ and $w = 100$ nm, the Hoffman model predicts a critical island gap of 20 Å and coalescence stress of 2 GPa. However from stress measurements during deposition of the noble metals Ag, Cu, and Au, the maximum tensile stresses rarely exceed 100 MPa.[4,12,13] Even for refractory materials such as Ti and Cr that grow with much smaller island sizes (e.g., $w = 5$ nm yields a predicted stress of almost 9 GPa), the maximum measured tensile stress is only about 1 GPa.[14,15]

For comparison with results later in the article, a uniaxial strain geometry is also considered and modeled as a semi-infinite slab of width $w$ as shown in Fig. 1(b). The slab configuration is more amenable to molecular dynamics calculations because there are no edges or corners (which tend to round off or facet), and thus will allow more direct comparisons between analytical and simulation results. In this configuration, the coalescence strain is uniaxial along the x-direction and zero in the orthogonal directions. A similar energy balance calculation, as described above, for the geometry depicted in Fig. 1(b) results in exactly the same expressions for the critical gap in Eq. (2) and coalescence stress in Eq. (3), except that the biaxial modulus is replaced by $M = E(1-\nu)/(1-\nu-2\nu^2)$, which equals 99 GPa using $E = 67$ GPa and $\nu = 0.33$. This slab geometry is assumed in further model calculations unless otherwise indicated.

It should be noted that NC implemented the Hoffman model to derive an upper bound estimate of the magnitude of coalescence stresses and to motivate an alternative model of coalescence stress generation via grain boundary zipping of elliptical grains via a crack-closure mechanism[9]. In addition, they recognized that the energetic balance argument in the Hoffman model was limited in that is does not predict how much smaller the gap between islands must become for coalescence to occur. However the upper bound estimate derived by NC is often quoted in the literature despite the fact that it is considerably larger than experimental observations.

In this article, we follow the same underlying mechanism suggested by Hoffman and NC but consider the energetics of coalescence *during* gap closure. We first examine how the interfacial energy between two closely spaced metallic surfaces varies with separation based upon molecular statics calculations using embedded atom method potentials. The interfacial energy plus strain energy provides a measure of the total system energy during the coalescence process. We will show that depending on the initial spacing between island surfaces, coalescence can be thermodynamically favored. In these cases where the energy of the coalesced islands is lower in energy than the separated islands, gap closure can occur either spontaneously or be kinetically limited due to an energetic barrier. In addition, we examine via molecular dynamics simulations of island coalescence to determine how large of an energetic barrier can be overcome at room temperature and compare our result with the Hoffman model.

**II. ENERGETIC ANALYSIS OF THE ISLAND COALESCENCE PROCESS**

The NC interpretation of the Hoffman model derived from the energy balance arguments are thermodynamic, and not kinetic, in origin. Coalescence is assumed to occur if the energy of the final state consisting of a single joined interface is lower than the initial state of two separated surfaces. The Hoffman model predicts that coalescence can take place between island separated by more than 50 Å (for $w$= 500 nm). Since this distance is much greater than the range of atomic interactions for metals, the model implies coalescence can occur for two surfaces that are essentially non-interacting. Improvements in the Hoffman model can be made by considering not just the energy of the final and initial states, but the energy of the system as it transitions between the two state points. In other words, we wish to examine the energetics during the coalescence process. We begin by examining how the energy between two closely spaced metallic surfaces varies as a function of separation, where the elimination of these free surfaces is the driving force for the coalescence process.

**A. Interfacial energy vs separation between closely spaced surfaces**

When closely spaced metallic surfaces approach each other, the nature of their metallic bonds is such that an attractive force develops between them. Therefore the interfacial energy between these two surfaces must also vary with separation. *Ab initio* calculations of interfacial energy vs separation support the theory of a universal binding energy relation (UBER) that appears to be valid for range of metallic and even covalent materials.[16-18] The UBER has been applied to problems ranging from adhesive avalanche issues in scanning probe microscopy[19,20] to crack propagation.[21]

In this method, unrelaxed rigid surfaces are brought incrementally closer together, starting from an initially large separation, and the change in energy is calculated at each separation. The resulting excess energy density, which we will refer to as the interfacial energy $\gamma_i$, vs separation $\delta$ for two neighboring surfaces can be well fit to a Rydberg function:

$$\gamma_i(\delta) = 2\gamma_s - (2\gamma_s - \gamma_0)\left[1 + \frac{\delta}{\delta_{infl}}\right]\exp\left[-\frac{\delta}{\delta_{infl}}\right], \tag{4}$$

where $\gamma_s$ is the unrelaxed surface energy at infinite separation, $\gamma_0$ is the interfacial energy at $\delta= 0$ (if the neighboring surfaces have different crystallographic orientations then $\gamma_0$ is the grain boundary energy), and $\delta_{infl}$ is the inflection point of the interfacial energy curve. Since no relaxation is allowed in the bulk or surface, $\delta= 0$ is defined as the separation at which the distance between surfaces equals the equilibrium interplanar spacing. The magnitude of $\delta_{infl}$ is related to the Thomas-Fermi screening length.[16] The derivative of $\gamma_i(\delta)$ is the traction $T$ acting on the surface (expressed as a force per unit area) due to the presence of a nearby surface and reaches a maximum value at $\delta_{infl}$. Our goal is fit the Rydberg function to interfacial energies vs separation calculated from molecular statics using interatomic potentials. The Rydberg function fit provides an analytical form for the interfacial energies that will be used for analytical solutions related to island coalescence.

The embedded atom method (EAM)[22,23] is a widely accepted technique for describing the interatomic potentials for metals. Conventional EAM potentials describing transition metals[24,25] such as Au, Cu, and Ni have cutoff radii $r_{cut}$ (i.e., the maximum distance between atoms included in the calculation of interatomic interactions) that include up to 3$^{rd}$ nearest neighbors such that typically $r_{cut} \approx 5$ Å. Because coalescence gaps are expected to be larger than 5 Å, we have created, using the general method of Voter and Chen,[25] potentials with longer cutoffs ($r_{cut}= 7.5$ Å and $r_{cut}= 15$ Å). Although the potentials are based loosely upon the properties of Au, in the results to follow we will refer to the material as "EAM Metal". It should be stressed that our goal is not to create an accurate potential for Au, but to be able to study the influence of a larger $r_{cut}$ on surface interactions and coalescence phenomena while maintaining reasonable material properties. The properties for our EAM Metal potential with $r_{cut}= 7.5$ Å are summarized in Table I (the properties with $r_{cut}= 15$ Å are very similar). Following the UBER methodology, the calculated interfacial energy vs separation $\gamma_i(\delta)$ for EAM Metal with $r_{cut}= 7.5$ Å and (100) surface normals at 0 K is shown in Fig. 2(a). The associated surface traction vs separation $T(\delta)$ reaches a maximum value of 22.35 GPa, as shown in Fig. 2(b). The interfacial energy and surface traction data are well fit by Eq. (4) with $\delta_{infl}= 0.52$ Å.

Note that $\gamma_0= 0$ at $\delta= 0$ because the interface that forms between the opposing (100) surfaces is fully coherent. We find that $\delta_{infl}$ does not change appreciably for potentials with longer cutoffs (e.g., $\delta_{infl}= 0.53$ Å for $r_{cut}= 15$ Å). It should be noted that EAM potentials are fit to the cohesive energy as a function of lattice constant via the universal binding curve.[22] The universal binding curve is the foundation on which Rose proposed the UBER to describe interfacial energies vs separation.[16] Therefore, the dependence of the interfacial energy on separation calculated using the EAM potential should be fairly realistic for metal systems. Since the UBER has been shown to accurately predict interfacial energies for covalent materials, we propose that this technique could be applied to refractory materials as well.

### B. Thermodynamic analysis of slab coalescence

Now that the interfacial energy is known as functions of separation, we can examine the energetic landscape during coalescence. For the slab geometry in Fig. 1(b) and as a consequence of the periodic boundary conditions, coalescence of a one dimensional array of islands is equivalent to coalescence between the two free surfaces of a single slab. For a slab at an initial separation $\alpha$ and under zero initial stress, the increase in strain energy per unit area due to stretching the slab to a closer separation $\delta$ is $\Delta E_\varepsilon = \frac{1}{2} Mw[(\alpha-\delta)/w]^2$, and the corresponding decrease in interfacial energy is $\Delta E_\gamma = \gamma_i(\delta) - \gamma_i(\alpha)$, where $\gamma_i$ is described by the Rydberg function of Eq. (4). Therefore the total change in energy $\Delta E = \Delta E_\varepsilon + \Delta E_\gamma$ for an initially unstrained slab (i.e., $\varepsilon= 0$) as a function of separation during coalescence is:

$$\Delta E_{\varepsilon=0}(\delta) = \frac{1}{2} Mw \left(\frac{\alpha-\delta}{w}\right)^2 + 2\gamma_s \left[\left[1+\frac{\delta}{\delta_{infl}}\right]\exp\left[-\frac{\delta}{\delta_{infl}}\right] - \left[1+\frac{\alpha}{\delta_{infl}}\right]\exp\left[-\frac{\alpha}{\delta_{infl}}\right]\right], \quad (5)$$

where it is assumed that a coherent boundary is formed at $\delta= 0$ so that $\gamma_0= 0$.

Ideally, we would now solve Eq. (5) analytically to determine the critical initial spacings for the following two limiting cases: (1) $\Delta E_{\varepsilon=0}(\delta=0) = 0$, which corresponds to the energy balance solution similar to the Hoffman model in that the final coalesced state at $\delta= 0$ has the same energy as the starting condition at $\delta= \alpha$ and, (2) $d(\Delta E_{\varepsilon=0})/d\delta < 0$ over $(0 < \delta < \alpha)$, which corresponds to spontaneous coalescence since the process of closing the gap is energetically favored at all times. Unfortunately, analytical solutions for these critical initial separations are difficult to obtain because of the linear-exponential nature of Eq. (5). Instead, we will obtain numerical solutions to further explore the energetics of the coalescence process. As will be shown, Eq. (5) does not fully capture the energetics of the problem because of the assumption of an initially unstrained slab. Modifications must be made to include the strain energy contribution due to stresses that result from interfacial forces *prior* to coalescence.

**C. Kinetically limited coalescence**

Consider an unstrained 10 nm-wide slab of EAM Metal with some initial separation $\alpha$ between the surfaces. Using Eq. (5) and the materials properties shown in Table I, we can numerically solve for the initial spacing for which $\Delta E_{\varepsilon=0}(\delta=0) = 0$ and find that $\alpha = 5.52$ Å. The change in the combined energy $\Delta E_\varepsilon + \Delta E_\gamma$ as a function of separation for a gap of 5.52 Å is shown in Fig. 3(a). Here, the energy of the fully coalesced slab is equal to the starting energy, i.e., coalescence is thermodynamically favored. The result is nearly identical to the Hoffman prediction ($\alpha_{crit} = 5.54$ Å from Eq. (2)). However as can be seen in Fig. 3(a), a large barrier exists in the energetic pathway to fully close the gap and therefore coalescence is kinetically limited. As discussed later in the article, energetic barriers to coalescence can potentially be overcome at finite temperature due to thermal fluctuations, where the magnitude of the fluctuations will dictate how large of an energetic barrier can be breached.

Closer inspection of the energy vs separation curve in Fig. 3(a) reveals two additional regions of interest. At nearly the initial separation (see inset), it is energetically favorable for the slab to stretch to a slightly closer separation resulting in a stress *prior* to coalescence (assuming the large energetic barrier is not overcome). By taking the derivative of $\Delta E_\varepsilon + \Delta E_\gamma$ and determining the position of the shallow local minimum, this pre-coalescence stress $\sigma_{pre}$ is found to equal the surface traction $T$ (see Fig. 2(b)) evaluated at the position of the local minimum ($\delta$ slightly less than $\alpha$) and to be independent of the slab width:

$$\sigma_{pre} = 2\gamma_s \frac{\delta}{\delta_{infl}^2} \exp\left[-\frac{\delta}{\delta_{infl}}\right]. \tag{6}$$

This result follows intuitively from the observation that the traction $T$ represents a force acting over the area of the interface, which is by definition a stress. If an increment of growth moves the now pre-stressed slabs closer together, $\sigma_{pre}$ will continue to exactly equal $T$ provided that the increments of growth are infinitesimal. In reality the spacing between islands changes in discrete atomic-spacing increments, however a continuum description of growth is convenient and still accurately represents the relevant phenomena. By including the strain energy contribution of the pre-coalescence stress from Eq. (6) with the previous expression in Eq. (5) for the energy change of an unstrained system, we arrive at the total change in energy per unit area during coalescence of a slab:

$$\Delta E(\delta) = \frac{1}{2} M w \left( \frac{\alpha - \delta}{w} \right)^2 + 2 \gamma_s \left[ \left[ 1 + \frac{\delta}{\delta_{nfl}} \right] \exp\left[ -\frac{\delta}{\delta_{nfl}} \right] - \left[ 1 + \frac{\alpha(\alpha - (\delta - \delta_{nfl}))}{\delta_{nfl}^2} \right] \exp\left[ -\frac{\alpha}{\delta_{nfl}} \right] \right] .$$

(7)

If we now use Eq. (7) to solve for the initial spacing for which $\Delta E(\delta = 0) = 0$, we find that $\alpha = 5.51$ Å for the 10 nm-wide EAM Metal slab. The solution is only fraction of an Å smaller than that from Eq. (5) because $\sigma_{pre}$ is very small (~ 15 MPa) at this relatively large separation.

The second point of interest in Fig. 3(a) occurs at nearly zero separation ($\delta = $ ~0.1 Å) where there exists an energy minimum. The driving force for coalescence, which is the decrease in interfacial energy with decreasing separation, approaches zero as $\delta$ nears zero as can be seen in Fig. 2(a). In contrast, the change in strain energy increases linearly with decreasing separation (i.e., the derivative of strain energy is linear with separation). Therefore the calculated change in total energy during coalescence will always exhibit a minimum near zero separation. Later in the article, we examine if this calculated minimum near zero separation is observed during atomistic simulations of coalescence and discuss the origins of any discrepancies between the simulated and calculated results.

**D. Spontaneous coalescence**

As the separation between slabs continues to decrease due to growth, the magnitude of the energetic barrier also decreases until finally at some critical separation the process has zero barrier and can occur spontaneously. This critical separation for spontaneous coalescence $\alpha_{spont}$ can be found numerically using Eq. (7) by determining the largest separation for which $d(\Delta E_{\varepsilon = 0})/d\delta < 0$ over (0.1 Å $< \delta < \alpha$). As shown in Fig. 3(b) for a 10 nm-wide EAM Metal slab, the critical separation for spontaneous coalescence is 2.89 Å. The inset in Fig. 3(b) shows that there is neither a shallow local minimum near the initial separation nor an energetic barrier to coalescence; hence, coalescence can proceed energetically "downhill" (ignoring the shallow minimum near $\delta = $ ~0.1 Å). The total stress resulting from spontaneous coalescence is therefore $M(\alpha_{spont}/w) + \sigma_{pre}$, where the pre-coalescence stress is given by Eq. (6) and evaluated at $\delta = \alpha_{spont}$.

For EAM Metal slabs with (100) surfaces and material properties listed in Table I, the coalescence gap vs slab width was calculated for the kinetically limited and spontaneous coalescence solutions, as shown in Fig. 4(a), while the corresponding coalescence stress vs slab width is shown in Fig. 4(b). As mentioned previously, the kinetically limited solutions yield almost identical results when compared to the Hoffman model, as indicated in the legends of the plots in Fig. 4. Note that the spontaneous coalescence model displays a stronger size dependence to the coalescence stress ($w^{-0.9}$) than the $w^{-0.5}$ dependence of the kinetically limited (Hoffman) model. The stronger $w$ dependence for the spontaneous coalescence model is a consequence of its weaker $w$ dependence on coalescence gap since to first order the coalescence stress goes as $\alpha/w$ (where $\alpha \propto w^n$ with $n < 1$). The weak $w$ dependence to the spontaneous coalescence gap is a consequence of the limited range of interaction between surfaces (see Fig. 2(b)) that drives the coalescence process.

To test the predictions of the analytical models, we have performed atomistic simulations of slab coalescence using conjugate gradient energy minimization (CGEM) calculations. EAM Metal slabs with widths $w$ ranging from 5 to 100 nm are created with (100) free surfaces. The directions orthogonal to the width direction (i.e., in the y and z directions) are periodic in order to emulate a semi-infinite slab, as in Fig. 1(b). Varying the lateral dimensions of the slab $l$ (see Fig. 1(b)) did not have any influence on the CGEM results. After equilibration at very large separation, the free surfaces are brought to just within the cutoff distance of the potential by decreasing the simulation box dimension along the periodic x-direction. Subsequently, the surfaces are moved closer together in 0.01 Å increments and CGEM is performed until the energy converges to within $10^{-6}$ eV tolerance of the total energy. The slab separation and the volume averaged virial stress are recorded after each step until coalescence occurs. The resulting approach curve (i.e., stress vs separation) for a 10 nm-wide EAM Metal slab is shown in Fig. 5, along with a comparison of the predictions from the spontaneous coalescence model. The stress prior to gap closure (from 5.5 to about 2.9 Å), which increases as the separation is narrowed, is the pre-coalescence stress and reaches a maximum value of 1.1 GPa. The discontinuous jump in stress occurs when the spontaneous coalescence gap is reached and results in a final stress of 6.5 GPa in the CGEM simulation of coalescence.

The CGEM simulation results for several different slab widths are overlaid with the model predictions in Fig. 4. The simulation results using the EAM Metal potential with $r_{cut}$= 15 Å produce nearly identical results. Because the spontaneous coalescence model requires that gap closure proceeds energetically downhill, the simulations are expected to give very similar results because of the nature of the CGEM scheme. The slight discrepancy in coalescence stress between the spontaneous coalescence model and the CGEM results, especially for small slab widths, is primarily due to non-linear elastic behavior of the EAM Metal potential. For example, the spontaneous coalescence gap for the 10 nm slab is 2.9 Å, which corresponds to a strain of 2.9%. At 2.9% strain, the modulus $M= E(1-\nu)/(1-\nu-2\nu^2)$ for the EAM Metal potential is softened by 10% compared to the modulus in the small strain limit. The stresses from the spontaneous coalescence model are calculated using the small strain limit modulus and therefore overestimate the coalescence stress for the 10 nm slab by about 10%, as can be seen in both Figs. 4 and 5.

Any remaining differences between the spontaneous coalescence model and the CGEM simulation results are likely due to the use of the UBER function to describe the interfacial energy in the model. In the molecular statics calculation of the interfacial energy in Fig. 2(a), the surfaces are assumed to be bulk-terminated and are not allowed to relax (or else coalescence would occur). In the CGEM simulations at large separations, the free surfaces can relax and experience a ~0.1 Å inward contraction[24] resulting in a slight decrease in surface energy. As the coalescing surfaces move closer together, the surface atoms slightly adjust their positions as they begin to interact with the adjacent surface. In addition, the UBER calculation does not account for any effects that the pre-coalescence stress could have on the interfacial energy. Although we do not take surface relaxation or stress into account in our interfacial energy vs separation calculations, more detailed UBER treatments have been considered in the literature.[27] Finally, it should be noted that the CGEM simulations do not show any indications of the shallow minimum near zero separation, as shown in the model calculations in Fig. 3, which may indicate that more careful interfacial energy calculations are required. However, we feel that these slight discrepancies between model and simulation do not significantly alter the conclusions.

**E. Thermally activated coalescence due to thermal fluctuations**

The position of a free surface at finite temperature will fluctuate over time resulting in a varying separation between opposing slab surfaces. These thermal fluctuations can potentially provide the activation energy necessary to overcome the energetic barrier to coalescence. However, the magnitude and temporal/spatial frequency of these fluctuations along with their dependencies on temperature and system size are not known. Because of the expectation of long run times (several ns) and relatively large dimensions (up to 100 nm), we restrict our molecular dynamics (MD) calculations to two-dimensions (2D) using the same EAM Metal potential. As a consequence of using a 2D system, the properties of the EAM Metal at 300 K change significantly as shown in Table I. The NC solutions for the coalescence gap and stress from the Hoffman model, given by Eqs. (2) and (3) respectively, remain the same except that the modulus for the 2D solutions is $M= E/(1-v^2)$. Even with these differences, the general conclusions to be drawn from this analysis are still comparable to the results already shown in the article.

Two-dimensional EAM Metal slabs are created with widths $w$ ranging from 5 to 100 nm and lateral dimension (i.e., in the y-direction) of either $l= 5$ nm or $l= w$, where coalescence occurs along the x-direction analogous to Fig. 1(b). The lateral dimension $l$ was varied to determine the surface fluctuation dependence on system size. At a separation greater than the cutoff of the potential, the different size systems are run for 10 ns (i.e., $10^7$ timesteps of 1 fs) under constant NVT integration at 300 K using a Nose-Hoover thermostat. The positions of all surface atoms are recorded every 0.5 ps and used to calculate the local slab width as a function of the distance $y$ along the interface. From $w(y,t)$ the maximum local width $w_{max}(t)$ can be determined for that time step. We are interested in the maximum local width because we propose that these local perturbations are the regions where coalescence will initiate. For example, $w_{max}(t) - w$ for an EAM

Metal slab at 300 K with $w$= 20 nm and $l$= 5 nm is shown in Fig. 6(a), along with the average stress in the width direction $\sigma_x$. Note that the nominal slab width is $w$= 20 nm but the actual time-averaged slab width is $w$= 19.901 nm. Of note in Fig. 6(a) is the strong correlation between the variations in the slab width and the stress. The period of the fluctuations $t_{fluc}$= 9.5 ps for both is almost an order of magnitude greater than the temperature oscillations from the thermostat (~1 ps) so stress and temperature do not appear to be correlated. In addition, the period of the fluctuations is independent of the NVT thermostat time constant and is no different if run under constant NVE conditions. So the variations in slab width are a result of elastic deformations from thermal phonons rather than bulk thermal expansion due to temperature variations.

Provided sufficiently long simulations are performed, $w_{max}(t) - w$ is well fit by a Gaussian distribution. For the same EAM Metal slab ($w$= 20 nm and $l$= 5 nm), the Gaussian fit to $w_{max}(t) - w$ in Fig. 6(b) yields a mean $\mu$= 0.24 Å and standard deviation $s$= 0.17 Å. Note that $\mu$ does not equal zero because we are examining the maximum slab thickness which will always be greater than the average slab thickness. A statistically significant perturbation in the slab thickness is therefore $w_{fluc} = \mu + 3\,s$, which should capture 99.74% of observed events. Table II is a compilation of $w_{fluc}$ and $t_{fluc}$ for all slab geometries run for 10 ns at 300K. Somewhat surprisingly, $w_{fluc}$ does not change significantly with $l$ for constant $w$. The systems with the larger $l$ should support modes with longer wavelengths and therefore larger amplitudes, but our results indicate that these modes are not sampled even with the relatively long MD timescales. However, the trend of increasing $w_{fluc}$ with increasing $w$ clearly indicates that larger fluctuation exist for the wider slabs. Finally, we recognize that MD simulations will always predict a conservatively small value for the maximum width fluctuation because of the limited timescale (in this case, 10ns) but the excellent Gaussian fit to the data indicates that significantly larger fluctuations are unlikely.

Now that we have an estimate of the magnitude of slab separation fluctuations $w_{fluc}$, we can examine how large of an energetic barrier to coalescence can be overcome. Because the results in Table II are statistical in nature, we choose to simply round off the values so that the following ($w$, $w_{fluc}$) pairs are assumed independent of $l$: (10 nm, 0.6 Å), (20 nm, 0.8 Å), (50 nm, 1.1 Å), (100 nm, 1.6 Å). As a reminder, these fluctuations are local perturbations and do not represent the entire slab surface achieving a closer separation. However, we will assume as much in order to use Eq. (7) to calculate the energetic barrier to coalescence as a function of initial separation. By making this allowance, we are presuming that a small region of a larger surface can locally coalesce based upon the same energetic analysis without significant error. An additional assumption is that once a local region coalesces, it will proceed laterally resulting in gap closure across the entire surface. Later in the article, we examine MD simulations of coalescing slabs with varying dimensions to try to validate these assumptions.

For a 2D 10 nm-wide slab of EAM Metal at 300 K, the energy as a function of separation for an initial gap of 2.85 Å as calculated using Eq. (7) is shown in Fig. 7. The distance necessary to crest the energetic barrier is about 0.6 Å, which is approximately equal to $w_{fluc}$ for a 10 nm slab as determined from the MD calculations. Therefore, the 10 nm slab is predicted to close a 2.85 Å gap based upon this thermally activated coalescence model, which is larger than the calculated spontaneous coalescence gap of 2.53 Å. However, this gap is still much smaller than the kinetically limited solution of 4.19 Å. Also note that the energetic barrier shown in Fig. 7 is 20 times smaller than the barrier that exists for the kinetically limited solution (i.e., Hoffman model). Similar thermally activated coalescence solutions are calculated for the other slab widths using the values of $w_{fluc}$ determined previously, and are compared to the spontaneous coalescence and kinetically limited results in Fig 8.

To compare against the predictions from the thermally activated coalescence model, MD simulations of slab coalescence for 2D EAM Metal at 300 K are performed for the slab geometries listed in Table II. The free surfaces of a slab are brought closer together in 0.01 Å increments and allowed to anneal for 1 ns (i.e., $10^6$ timesteps of 1 fs) after each change in separation. The slab separation and volume averaged virial stress are recorded every 0.5 ps to determine when coalescence occurs. The MD simulation results for all slab geometries are overlaid with the model predictions in Fig. 8. From the MD simulations, coalescence for a given $w$ is independent of $l$ so only one set of MD results is shown in Fig. 8. Upon closer inspection of the simulation results for which $l= w$, coalescence occurs as a result of a two-step process. A small, stable perturbation forms that eventually results in local coalescence over a surface region of approximately 5 nm in lateral dimension. Gap closure then proceeds laterally along the remainder of the interface at a rate of approximately 1000 m/s, which is similar to the speed of crack propagation in metals.[28] The agreement between the MD results and the predictions from the thermally activated coalescence model supports the assumptions made in the energetic analysis. The small deviation in comparing the stresses at smaller slab widths is again due to non-linear elastic behavior of the EAM Metal potential. The relatively small difference between the MD simulation results and the spontaneous coalescence model indicates that 300 K provides only modest thermal activation and that only small energetic barriers can be surmounted. Therefore, the large energetic barrier present in the kinetically limited model cannot be easily overcome at modest temperatures and consequently the Hoffman model dramatically overestimates the magnitude of coalescence stresses.

Admittedly, the slab geometries assumed in this article are extremely idealized since actual discontinuous Volmer-Weber films consist of islands with hemispherical cap shapes and some degree of adhesion with the substrate. Therefore, our model of coalescing slabs represent an upper bound estimate of the coalescence stress because of the large planar interfacial area compared to a hemispherical island and lack of traction with a substrate. In addition, we have not considered any stress relief mechanisms that may mitigate the magnitude of the stresses resulting from coalescence. Consequently, further computational work assuming more realistic island shapes and varying degrees of adhesion with the substrate is being explored to determine how these features affect island coalescence and if inelastic phenomena are prevalent enough to modify the predictions herein.

## IV. CONCLUSION

We have analyzed island coalescence stress generation following an argument suggested by Hoffman that the mechanism is a trade off between strain energy generation due to the stretching of the islands and the energy decrease associated with the elimination of surface energy. However in contrast to a simple energy balance calculation, we have considered the total energy of the coalescing island during the entire gap closure process. The interfacial energy between two closely spaced metallic surfaces was calculated from molecular statics using embedded energy method potentials and shown to fit an analytical form derived in previous studies from *ab initio* calculations. We derived an analytical expression for the sum of the interfacial energy plus strain energy, given by Eq. (7), which allowed us to calculate the energy of impinging island during coalescence. In cases where coalescence was found to be thermodynamically favored, gap closure was found to occur either spontaneously or be kinetically limited due to an energetic barrier. Conjugate gradient energy minimization calculations of simulated coalescence agree extremely well with the predictions from the spontaneous coalescence model. Molecular dynamics simulations at room temperature demonstrate that thermal fluctuations can only reduce the local gap between impinging surface by about ~1 Å. By comparing these fluctuations in separation to the width of the energetic barrier to coalescence, we were able to calculate the expected coalescence gap and stress resulting from this thermally activated process. The relatively modest energetic barrier that could be overcome in the thermally activated coalescence process at room temperature helps explain why the Hoffman model overestimates the magnitude of the stress resulting from island coalescence.

## ACKNOWLEDGEMENTS

The authors would like to thank S M Foiles for many helpful discussions. Sandia National Laboratories is a multiprogram laboratory operated by Sandia Corporation, a Lockheed Martin Company, for the United States Department of Energy under Contract No. DE-AC04-94AL85000.

TABLE I. Material properties of EAM Metal with $r_{cut}$= 7.5 Å.

| Dimension – lattice struct.[a] | T [K] | E [GPa][b] | $\nu$ [][c] | $\gamma_s$ [J/m$^2$][d] |
|---|---|---|---|---|
| 3D – fcc | 0 | 96.42 | 0.4031 | 1.58 |
| 2D – hex | 300 | 489.32 | 0.4490 | 2.75 |

[a]The stable lattice phase is face-centered-cubic (fcc) in three-dimensions and hexagonal (hex) in two-dimensions.

[b]Young's modulus.

[c]Poisson ratio.

[d]Unrelaxed surface energy.

TABLE II. Slab width fluctuations from MD simulations of 2D EAM Metal at 300 K for 10 ns.

| $w$ [nm] | $l$ [nm] | $t_{fluc}$ [ps][a] | $w_{fluc}$ [Å][b] |
|---|---|---|---|
| 10 | 5 | 4.6 | 0.609 |
| 10 | 10 | 4.6 | 0.610 |
| 20 | 5 | 9.5 | 0.757 |
| 20 | 20 | 9.5 | 0.778 |
| 50 | 5 | 23.8 | 1.05 |
| 50 | 50 | 23.8 | 1.09 |
| 100 | 5 | 47.6 | 1.52 |
| 100 | 100 | 47.6 | 1.57 |

[a]Period of the surface oscillations.

[b]Slab width fluctuations are set equal to $\mu + 3\,s$ from Gaussian distribution fits to the probability distribution function of $w_{max}(t) - w$.

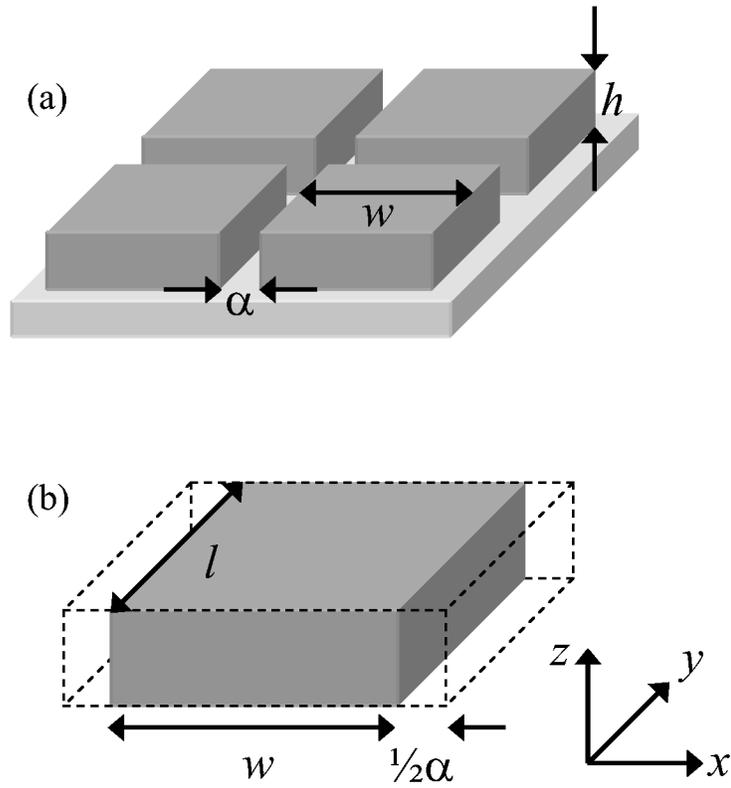

**FIG 1.** (a) Schematic diagram of a periodic array of square islands of thickness $h$ and width $w$ separated from each other by gap $\alpha$. (b) Schematic of an semi-infinite slab of width $w$ and lateral dimension $l$ with periodic boundaries indicated by dashed lines to represent a simplified island geometry.

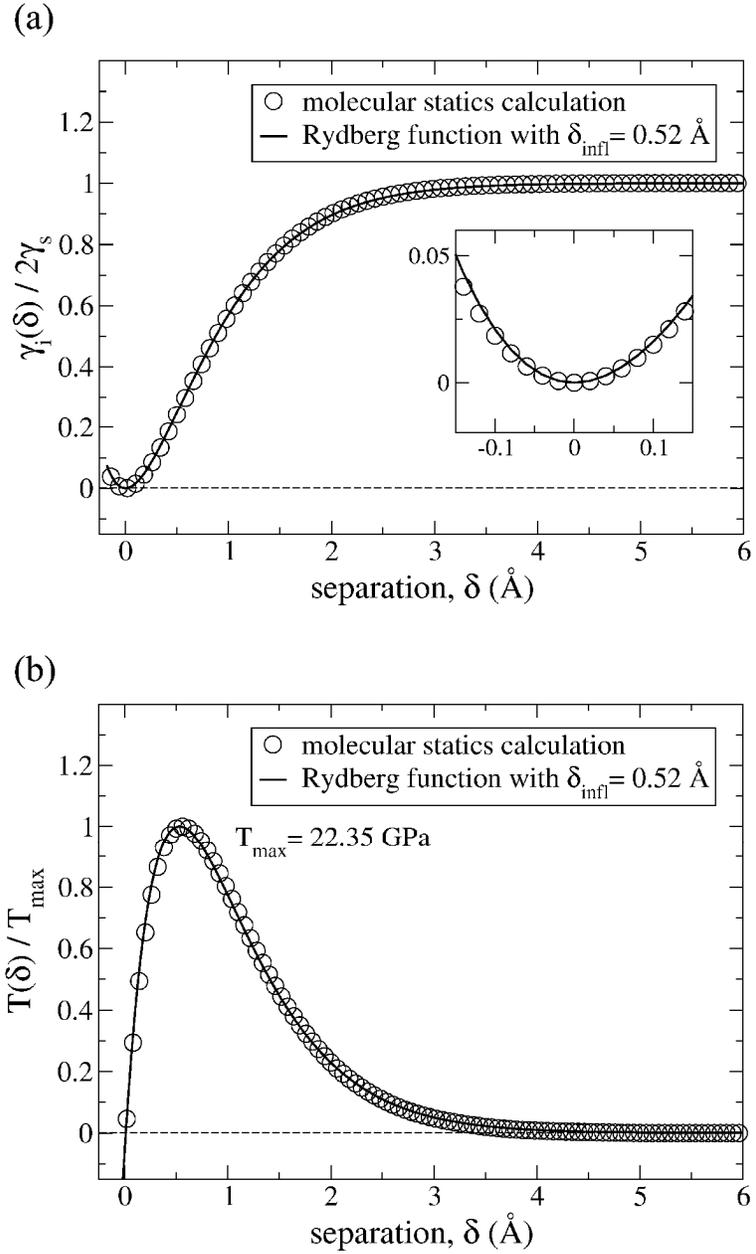

**FIG 2.** (a) Interfacial energy vs separation $\gamma_i(\delta)$ from molecular statics calculations of (100) surfaces of EAM Metal with $r_{cut}$= 7.5 Å and material properties listed in Table I. (b) The surface traction $T(\delta)$ equals the derivative of $\gamma_i(\delta)$ from (a). The fit of the Rydberg function described by Eq. (4) yields $\delta_{infl}$= 0.52 Å, as shown in both (a) and (b).

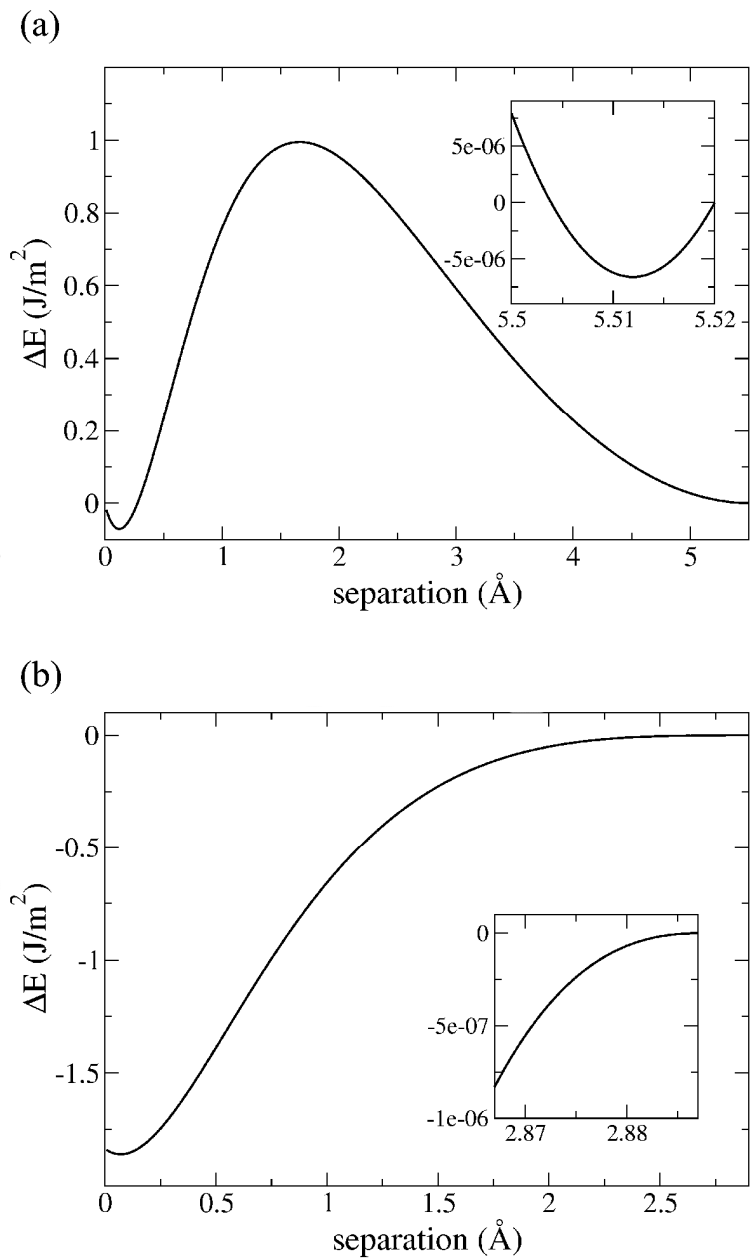

**FIG 3.** (a) Calculated change in energy by closing a gap of 5.52 Å between 10 nm-wide EAM Metal slabs. Inset: close up of energy for separations nearly equal to the initial gap. (b) Same slab geometry except that the initial gap is now only 2.89 Å.

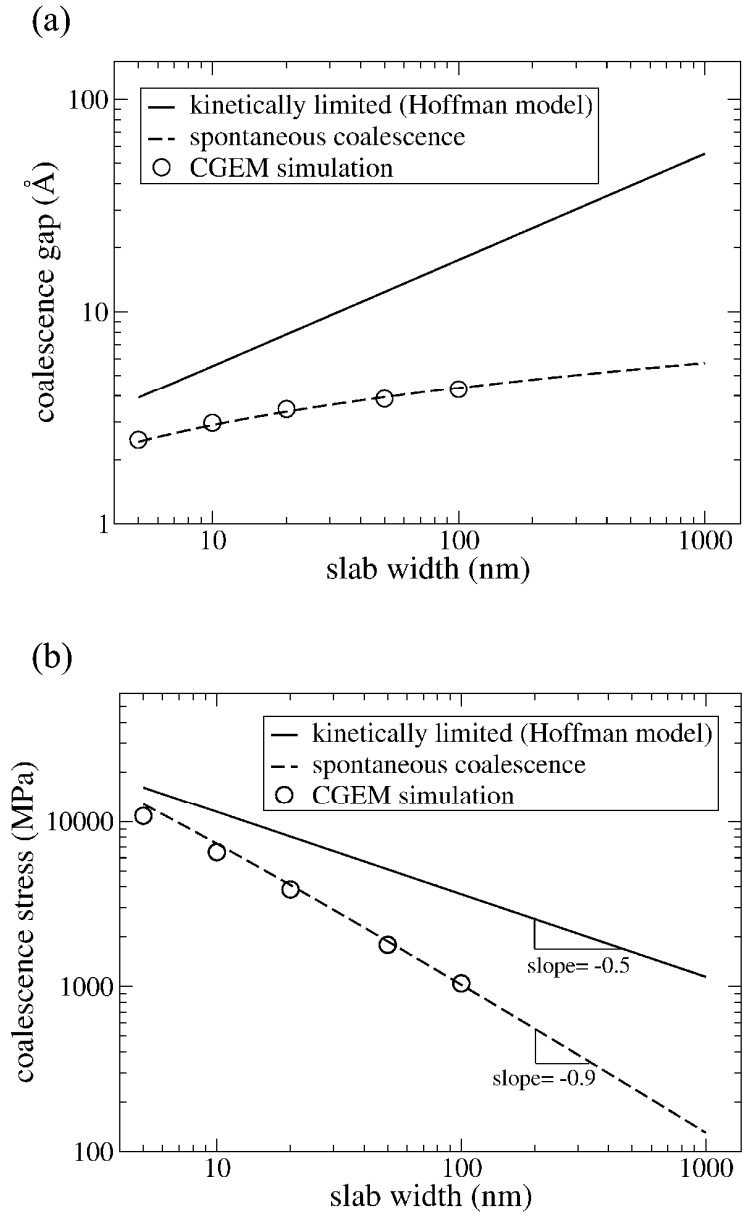

**FIG 4.** (a) Coalescence gap vs slab width comparing the kinetically limited solution (similar to the Hoffman model), the spontaneous coalescence model, and conjugate gradient energy minimization simulation results of slab coalescence for EAM Metal with (100) surfaces. (b) Coalescence stress vs slab width comparing the same models.

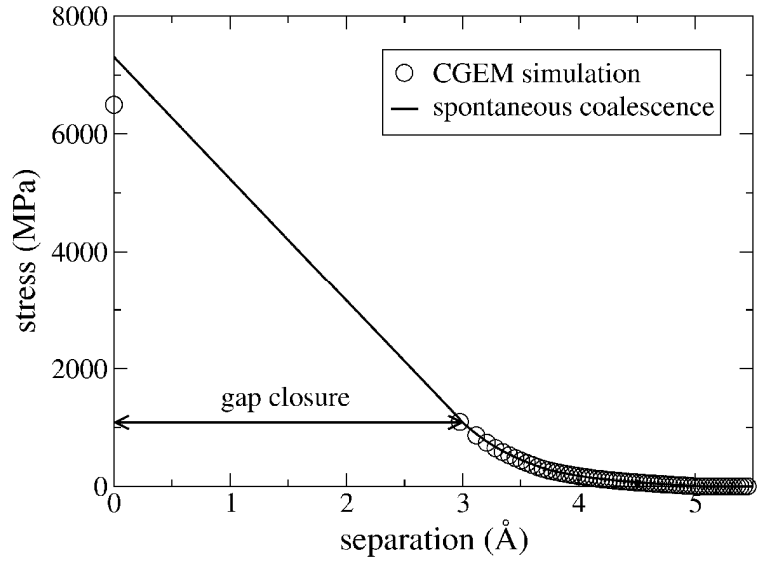

**FIG 5.** Approach curve for 10 nm-wide slab of EAM Metal as the (100) free surfaces get gradually closer together until coalescence occurs. The stress prior to gap closure is the pre-coalescence stress, given by Eq. 6, while the discontinuous jump in stress occurs when the spontaneous coalescence gap is reached.

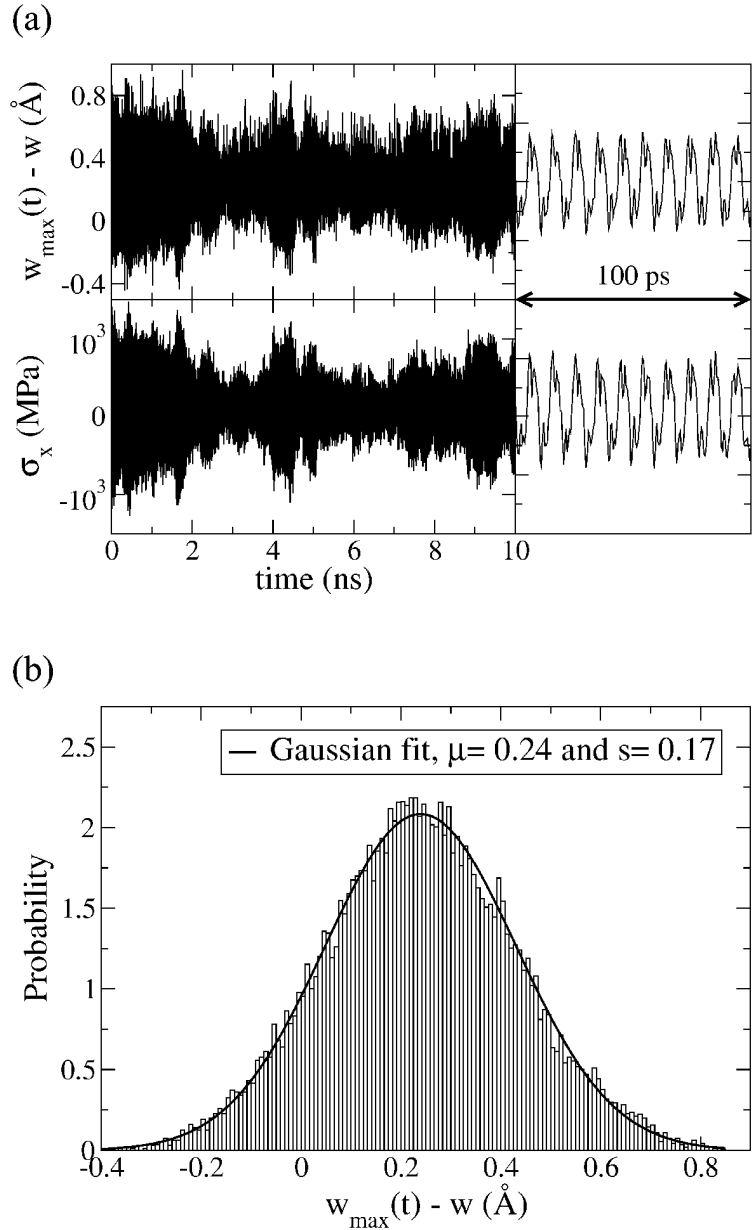

**FIG 6.** Molecular dynamics calculation under constant NVT integration at 300 K of a 2D EAM Metal slab ($w$= 20 nm and $l$= 5 nm) with (100) free surfaces. (a) Deviation in the maximum slab width $w_{max}$ and stress in the width direction $\sigma_x$ during the 10ns run and, (b) the probability distribution function of $w_{max}$.

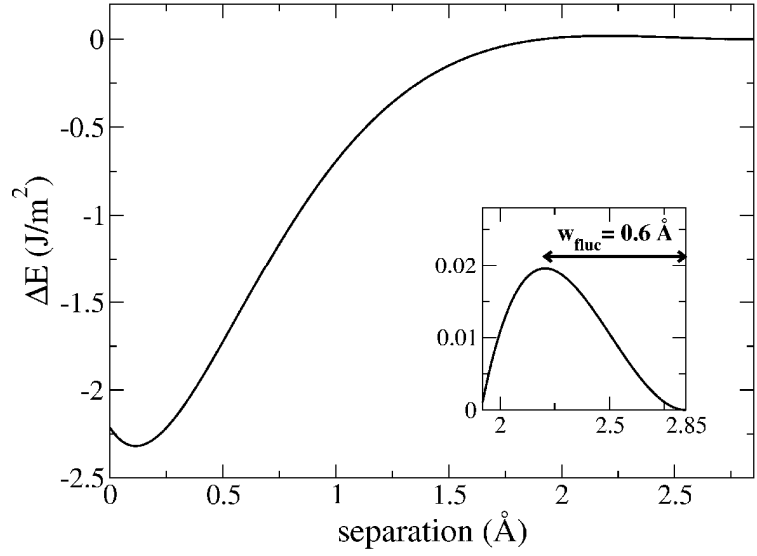

**FIG 7.** Calculated change in energy using Eq. (7) during closure of a 2.85 Å gap between 10 nm-wide slabs of 2D EAM Metal at 300K. Inset: close up showing the distance to the crest of the energetic barrier is 0.6 Å, which is equal to the magnitude of the maximum width fluctuation $w_{fluc}$ determined from MD calculations.

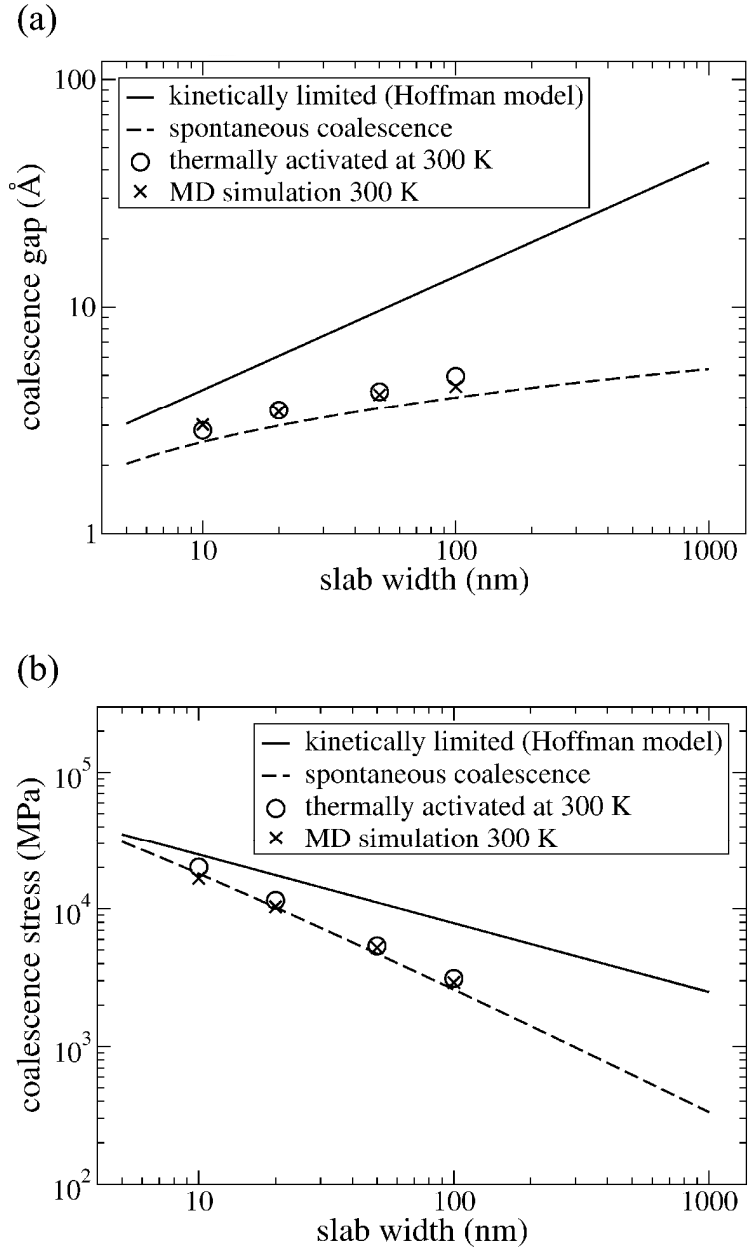

FIG 8. (a) Coalescence gap vs slab width for 2D slabs of EAM Metal with (100) surfaces at 300 K, comparing MD results of coalescing slabs with model predictions. (b) Coalescence stress vs slab width comparing the same models. Note that coalescence for a given width $w$ was found to be independent of the lateral dimension $l$ so only one set of MD simulation results is shown.